# Forensic Analysis of Instant Messenger Applications on Android Devices

Aditya Mahajan
Institute of Forensic Science
Gujarat Forensic Sciences
University, Gujarat, India

M. S. Dahiya
Institute of Forensic Science
Gujarat Forensic Sciences
University, Gujarat, India

H. P. Sanghvi
Directorate of Forensic Science
Gujarat State
Gujarat, India

## ABSTRACT
The modern day Smartphone's have built in apps like **"WhatsApp & Viber"** which allow users to exchange instant messages, share videos, audio's and images via Smartphone's instead of relying on their desktop Computers or laptop thereby increasing the portability and convenience for a layman smart phone user. An Instant Messenger (IM) can serve as a very useful yet very dangerous platform for the victim and the suspect to communicate. The increased use of Instant messengers on Android phones has turned to be the goldmine for mobile and computer forensic experts. Traces and Evidence left by applications can be held on Android phones and retrieving those potential evidences with right forensic technique is strongly required. This paper focuses on conducting forensic data analysis of 2 widely used IMs applications on Android phones: **WhatsApp and Viber**.

5 Android phones were analyzed covering 3 different versions of Android OS: Froyo (2.2), GingerBread (2.3.x) and Ice-Cream Sandwich (4.0.x). The tests and analysis were performed with the aim of determining what data and information can be found on the device's internal memory for instant messengers e.g. chat messaging logs and history, send & received image or video files, etc. Determining the location of data found from FileSystem Extraction of the device was also determined. The experiments and results show that heavy amount of potential evidences and valuable data can be found on Android phones by forensic investigators.

## General Terms
Android Forensics, Instant Messenger Forensics.

## Keywords
SmartPhone Forensics, Android Forensics, WhatsApp Forensics, Viber Forensics, Instant Messenger Forensics.

## 1. INTRODUCTION
There has been rapid increase in online communication in the last 7-8 years, especially in mobile communication. Smartphone's have taken up the market so well that everybody now can interact, socialize, and can share ideas and Information sitting at any corner in the world. Today's young generation is busy in chatting and messaging every time with friends and with unknowns too. People are continuously exchanging information like images, videos, activities and events. But despite of getting connected with friends for more and more time, their privacy [2012 Alfred] is also getting more vulnerable to threats by hackers and fraudsters. This is because criminals know that doing crimes using online mobile applications is secure as it is very tough task for extracting the information from mobile phone from which crime was committed. This is so because mobile phones have very less memory and that too they have flash memory which gets washed fast and easily on mobile phones. One more reason of using mobile applications for doing any crime is that their application logs will not get saved at Internet Service provider side. Criminals are also using mobile applications like **"WhatsApp"** and **"Viber"** because it's easy for them to exchange & share information before, while and after committing the crime. Table-1 Shows the applications with the information about its users and downloads.

**Table-1 Application and their users**

| Application | Information |
|---|---|
| WhatsApp | 100000000 + downloads (On Google Play) |
| Viber | 140 million people |

So Forensic Analysis of Instant Messenger applications [2010, Mohammad I. Husain] are very important from forensic point of view as it can play a crucial role in any cyber and crime investigations. Forensic Analysis will give details which will help investigators and investigation agencies in solving & relating the cases with the crime reported. Features of the 2 Instant Messenger Applications are listed in Table 2. This Paper is aimed to focus on forensic examination of data & Information stored by the application on the phone and the forensic data extraction techniques.

**Table-2 Application and their features**

| Application | Features |
|---|---|
| WhatsApp | 1. Text Chat<br>2. Send & Receive Images<br>3. Send & Receive Videos<br>4. Send & Receive Audio's<br>5. Group Chat<br>6. Sharing V-Cards & Contact Information |
| Viber | 1. Free Voice Calling<br>2. Free Texting |

## 2. FORENSIC CHALLENGES AND STORAGE ARTIFACTS

In any forensic examination of the Android phones, a forensically sound methodology has to be taken care at the most. The equipments, environment and techniques to be used should be as per the cardinal rules of computer forensics. A forensically sound methodology neither changes any data on the original device nor will it write any data on it.

### 2.1 Android Forensics & Challenges

This paper focuses on the forensic analysis of the data stored by two (2) Instant Messenger applications. But prior to examination of data, it is needed first to locate & extract the files and folders where the artifacts related to the applications have been stored in the internal memory of the phone. But forensic examination of applications and their databases is tough if it's encrypted [Android Encryption] or deleted. Also, Android phone users are mostly connected to Internet every time, so data can be wiped remotely by any person.

Also, Updates are continuously released by the developer of the application and Operating system installed in the phone which makes it hard for forensic examiners to understand every updated feature and to be ready to deal with new methods of forensic examination with available old tools only. Also, update creates more Challenges for Law Enforcers and forensic investigators to prove and provide the evidence in court of law [2006, Al-Zarouni, Marwan].

### 2.2 Proposed Method

The method for analyzing the data and information is very simple, however Cellebrite UFED (Universal Forensic Extraction Device) Classic Ultimate (V 1.8.0.0) is used for extracting the files and folders and is considered to be one of the best Hardware equipment in Digital Forensics Industry for mobile device extractions. Filesystem Extractions were carried out using UFED so as to understand the data stored in files and folders of the phone's internal memory. To understand the file and directory structure of storing the information in Android devices, 5 Android phones were taken with 3 different versions of Android OS which are listed in Table 3.

**Table-3 Android phones and their versions**

| Mobile Phone Make & Model | Android Version |
|---|---|
| Sony Xperia ST15i mini | 4.0.4 - ICS |
| Sony Xperia Neo V (MT11i) | 4.0.4 - ICS |
| LG P698 | 2.3.4 – Ginger Bread |
| Samsung GSM GT-S5830 Ace | 2.2.1 Froyo |
| HTC A8181 Desire | 2.2. - Froyo |

### 2.3 FileSystem Acquisition

In this stage, the FileSystem Extraction of every Android phone was performed. The extractions were performed using mobile acquisition Device "Cellebrite UFED (Universal Forensic Extraction Device) Classic Ultimate" under forensically sound environment. Before starting the acquisition process, "USB Debugging" option was "enabled" from settings menu of the phone. FileSystem extraction extracts all the folders and files stored in the Internal Memory of the Android phone which contains various files including Database Files, Configuration Files, etc. FileSystem extraction extracts data of every application in a separate folder stored on the phone. So for both the applications, FileSystem Extraction was done only once on every phone.

**Table 4 Application and their artifacts location**

| Application | Folder Location | Folder Name |
|---|---|---|
| WhatsApp | /data/data/com.whatsapp/ | /databases |
| Viber | /data/data/com.viber.voip | /databases |

**Table 5 Application and name of files containing artifacts**

| Application | File Name (with extension) | Table Name |
|---|---|---|
| WhatsApp | 1. Msgstore.db | 1.1 messages<br>1.2 chat_list |
|  | 2. wa.db | 2.1 wa_contacts<br>2.2 sqlite_sequence |
| Viber | 1. Viber_call_log.db | 1.1 Viber_call_log |
|  | 2. Viber_data (a database file) | 2.1 android_metadata<br>2.2 phonebook raw contact<br>2.3 phonebook contact<br>2.4 phonebook data<br>2.5 Viber numbers<br>2.6 Calls |
|  | 3. Viber_messages (a database file) | 3.1 android_metadata<br>3.2 messages'<br>3.3 sqlite_sequence<br>3.4 threads<br>3.5 participants |

### 2.4 Artifacts, Structure & Storage Locations

After FileSystem Extraction of Android Devices, many files and folders will be generated but it will be very difficult and time consuming for any forensic expert to analyze and look into every file of every folder. However, UFED physical Analyzer can also be used to view the information extracted from UFED but sometimes physical analyzer also gets fails in displaying information related to every application. This is described later in the paper. So, forensic investigators and experts should know the structure that how the forensic artifacts related to Instant Messengers are stored and to what location they are stored. So, Table 4 & 5 shows what artifacts are stored on which files with what extension.

Forensic examination of phones listed in Table-4 showed that a database folder and database files related to the "WhatsApp" application are stored on the phone's internal memory but however the pictures, videos and audio files downloaded by the application were found to be stored in the external memory card with folder name WhatsApp. A folder named "media" is stored in the memory card which contains all undeleted images, videos and audio files downloaded or uploaded using WhatsApp. However, the messages or Text Chat done by the user is stored in the database file located in internal memory of the phone. "WhatsApp" chat history logs were also found in the memory card but they were encrypted whereas the logs found in the phone's internal memory were stored in plain and readable text. Status, phone numbers, names and timestamps were also stored in the database folder in the internal memory of the phone. In case of "Viber" Application, only one "databases" named folder was found which was containing 5 database files out of which 3 database files plays very important role in forensic examination. The information stored in those 3 files are shown in Table-5. Similar to desktops and laptops in which Internet related artifacts can be found, Smartphone's like Android also stores data that can help in determining how the device and the applications installed in it were misused and what data can be important from forensic point of view.

## 3. Methodology

The main purpose of this research is to find that whether chatting and information sharing activities performed using Instant Messenger applications on Smartphone's are stored in phone's internal memory or not. If yes, then it will be very helpful for forensic experts in crime and cyber crime cases by examining those highly evidentiary values. The goal and aim of this research was achieved successfully.

FileSystem extractions were conducted on 5 Android phones and forensic examinations of 2 widely and commonly used Instant Messaging applications were performed: WhatsApp and Viber.

In real investigations of Android phones, the forensically sound methodology on extraction of data, time-stamps and generating Hash values plays a great value to prove the evidence in court of law. That is why all the experiments during this research were conducted using forensically accepted equipments under forensically accepted conditions and as per cardinal rules of computer forensics. It preserves the integrity of data on the phone, preventing any kind of contamination so as to get it accepted in court. **The Previous Version of the Cellebrite UFED v1.1.05 was tested by National Institute of Standards & Technology [NIST] in 2009[January 2009] to ensure the quality, reliability, validity of testing methods and results of examination, however here UFED v1.8.0.0 is used for examination[October 2012].** The experiments were conducted on Non-Rooted and rooted, both types of Android phones because in real time investigations, suspects may be using any one type of them.

### 3.1 Testing Equipments & Environment

For computer or mobile forensic examinations, forensic workstation, server, forensic Hardware's, software's and write blockers are necessary and during the analysis everything was set up. Table 3 shows the list of phones used for experiments and Table 6 shows the list of Hardware's, software's and applications used for the analysis.

### 3.2 Testing & Analysis

The testing process and Analysis was done in 3 stages
1. Scenario
2. FileSystem Analysis
3. Identification

### 3.2.1 Scenarios

The first stage consists of 2 scenarios:
1. Pre-Installed applications
2. Installing applications manually & conducting Activities

**Table-6 List of Hardware and software used for analysis (along with versions)**

| List of Hardware's & Software's Used for Experiments | Versions |
|---|---|
| Cellebrite UFED (Universal Forensic Extraction Device) Classic Ultimate | 1.8.0.0 |
| Cellebrite UFED Physical Analyzer (Software) | 3.6.1.6 |
| Android phones ( Versions- ICS, Gingerbread, Froyo) | Refer Table 3 |
| Sqlite Database Browser | 2.0 |
| WhatsApp & Viber applications for testing | Refer Table 3 |

**Pre-Installed Applications**- It is described as the stage when applications were already installed in the phone and were used heavily by the users before the research was conducted. The users were asked about the activities which they have performed while using the application before the research. It was found that "WhatsApp" application was heavily used by the users using almost all the features. However, "Viber" application was used less as compared to "WhatsApp". However all basic features of the "Viber" application were used by the users.

**Installing applications manually & conducting User Activities**- It is described as the stage when applications were not installed and used before the research, so they installed manually. The applications were downloaded from "Google Play Store" and then installed on the phones. Common user activities [2012, Noora Al Mutawa], were performed and all possible features of the applications were used. For every phone, a set of predefined activities performed using all the features of each application. Both the applications were continuously used for more than 2 weeks. Table-7 shows the Activities performed on each application of each Android phone.

**Table-7 Activities performed on applications**

| Application | Activities Performed |
|---|---|
| WhatsApp (WhatsApp to WhatsApp Users) | 1.Instant Messaging / Chatting<br>2.Sending Photos<br>3.Sending Videos<br>4.Sending Contacts<br>5.Sending Audio<br>6.Receiving Photos<br>7.Receiving Videos<br>8.Receiving Contacts<br>9.Receiving Audio |
| Viber (Viber to Viber ) | Online Voice Calls<br>Online SMS |

### 3.2.2 FileSystem Analysis

After performing the FileSystem Extraction, hash values were generated for every device. Hash values play very significant role in presenting the cyber case in any court of law [Kailash Kumar, 2012]. Hash values of every device are shown in Table 8. This last stage involves examination of the acquired files and directories by the FileSystem extractions of every Android phone. In this stage, it was to determine that whether

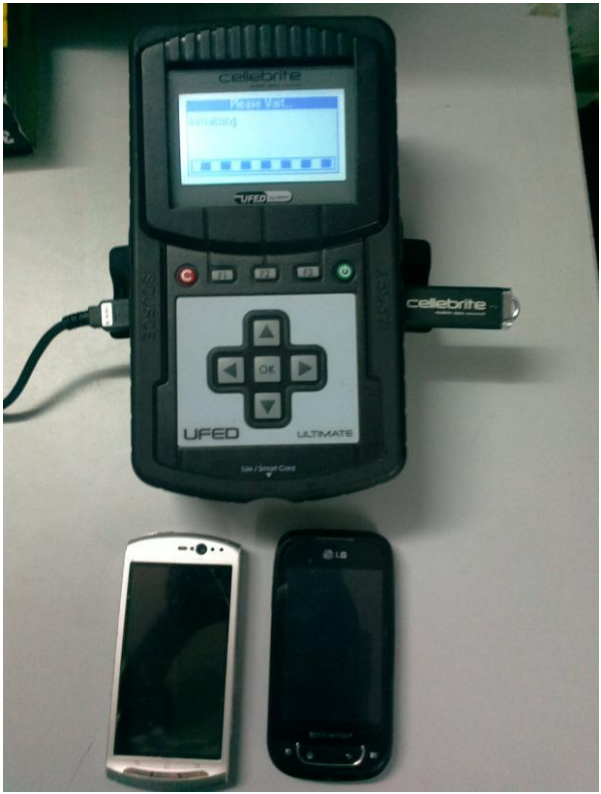
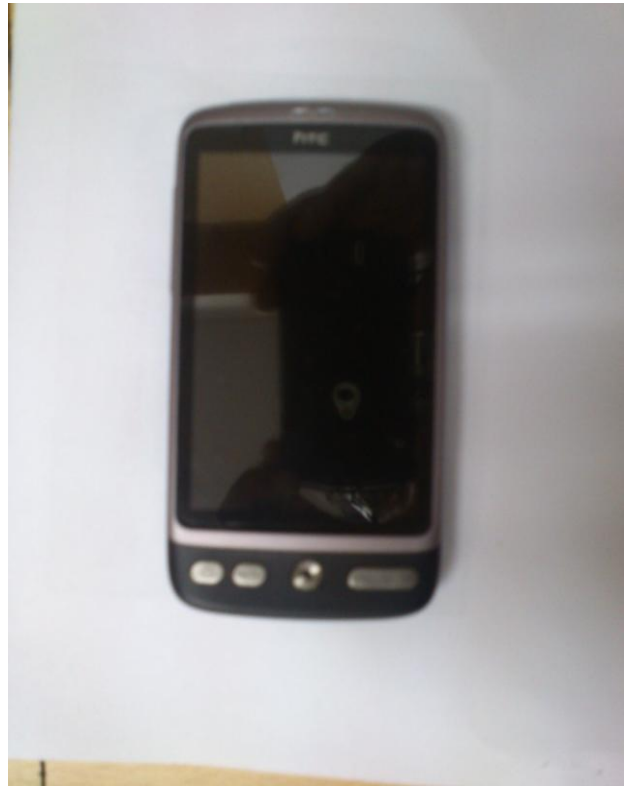

1. Cellebrite UFED
2. Sony Ericsson Xperia Neo V (MT11i)
3. LG-P698

4. HTC Desire A8181

**Picture 1: Equipment (UFED) and Android Phones used in research**

databases and logs generated by the application contain any relevant and important data or not. If any databases and logs are found, then the amount of data with its significance was determined using tools and techniques. Also, understanding the database structure and data related to Instant messaging applications was also determined. This stage is also described in 2 following stages:

1. **Analysis in UFED Physical Analyzer**

In this, the .ufd file generated by the UFED device was loaded in the Physical Analyzer software and then analysis was done for WhatsApp and Viber. The artifacts found in this software for "WhatsApp" application are shown later but No data related to "Viber" application was found in this software.

2. **Manual Analysis of data**

In Manual analysis, each folder and file which was extracted by UFED device was opened and analyzed. A folder named as "databases" was found in WhatsApp and Viber directories and several ".db" files were found containing important data.

### 3.2.3 Identification

This third stage involved identification of files and folders which are relevant to our research topic. In this step, it is determined whether application stores any logs of the activities performed by application in the internal memory of phone or not. Also, logs were searched in the internal memory of phone because logs were not stored on the external Memory Card of the phone. Each folder and directory was explored which was extracted using FileSystem extraction of the Android phone. As the Common User Activities and Extractions were performed on different versions of Android

**Table 8 – Hash Values generated by UFED for Android Devices**

| Android Device | Hash Value ( SHA 256 ) |
|---|---|
| Sony Xperia ST15i mini | B1AC9CEB516F2AF8543452EE4CE978A0BC9C9D5E466138B8869E13CCE2754299 |
| Sony Xperia Neo V (MT11i) | 6C3F63E2ECEB0C035CB4AE25B2083A661C2B79228A50DD92491D60AD4CC2B34 |
| LG P698 | A24D8BCCAE53A03195ED0AEB2AF90E34DBBA3B2741464B154812F7EC42CC68 |
| Samsung GSM GT-S5830 Ace | D79F70B761EF32545AF6D7EDF484DA2EEBD3C96B2D028BEE5EB413037076325F |
| HTC A8181 Desire | 23BD7BC18C36A20BC3F07501570A9EF5C787A6701C49E4E4A4EE80D6EFB761A6 |

**Table 9 – WhatsApp information for Physical Analyzer**

| | Artifacts Found | Artifacts Not Found |
|---|---|---|
| **"WhatsApp" Artifacts Related Information In Physical Analyzer** | Sent chats messages to every user | Contact list |
| | Received chat messages from every user | Profile picture of the User (If Any) |
| | Time Stamps of every chat session | Profile pictures of users with whom Chat Sessions were done |
| | | Location of downloaded images or videos via WhatsApp |

## 4.1 WhatsApp Application Forensic Examination & Artifacts

This section describes the forensic analysis of "WhatsApp" Instant Messenger application and Artifacts found during examination of the 5 phones with their results.

**Table-10 Artifacts found related to applications**

| | Activities Performed | Similar Data Forensic Examination (Found / Not Found ) | Artifact related to the "WhatsApp" account with which data was shared ( Found / Not Found) |
|---|---|---|---|
| **Artifacts Related Information while Manual Search of "WhatsApp" Application** | Login Phone Number of User | Not Found | N.A |
| | Received chat messages | Found with Timestamp | Found with Timestamp |
| | Outgoing Messages | Found with Timestamp | Found with Timestamp |
| | Send Images | Found with Timestamp. Sent Image File Name was also Found [ Location of photos was found to be in memory card] | Found with Timestamp |
| | Received Images | Found with Timestamp. Received Image File Name was also Found [ Location of photos was found to be in memory card] | Found with Timestamp |
| | Sent Videos | Found with Timestamp. Video File Name was also Found [ Location of videos was found to be in memory card] | Found with Timestamp |
| | Received Videos | Found with Timestamp. Video File Name was also Found [ Location of photos was found to be in memory card ] | Found with Timestamp |

phones, so it was to identify that whether directory structure differs or not with what version of Android Version is installed on the phone. Hence it was identified that directory structure remains same for different Android Versions. Also, it was identified that almost all User Activities with their logs were stored in the internal memory of the phone creating "databases" named folder for all activities along with date and time stamps. Detailed description is presented later in the paper.

## 4. EXPERIMENTS AND RESULTS

Both applications were came either pre-installed on the phone or they were first downloaded and them installed from Google Play Store. Out of 5 phones that were tested, three(3) phones were with pre-installed those 2 apps and the users were using those applications since more than 3 months and a big amount of information and data was retrieved from those phones which played a very important role for the accomplishment of this research. Other 2 phones, in which only "WhatsApp" was pre-installed and "Viber" was installed for conducting the research and several common user activities were performed which are already shown in Table-7.

### 4.1 WhatsApp Application Forensic Examination & Artifacts

This section describes the forensic analysis of "WhatsApp" Instant Messenger application and Artifacts found during examination of the 5 phones with their results.

#### 4.1.1 Artifacts Found in UFED Physical Analyzer

A ".ufd" file generated by the UFED device after FileSystem Extraction was generated along with a ".rar" file. ".ufd" file was opened in Physical Analyzer and it was found that the chat sessions done by the user using "WhatsApp" are present. Also, Timestamps were also present for every chat session done and Timestamps plays very important role. Table-9 shows the Artifacts related information.

#### 4.1.2 Artifacts found in Manual Analysis

It was found that "WhatsApp" application maintains and stores the logs and records of all the activities done by the user using the application including all that Chat or text messaging history in a database file named as "msgstore.db" and the contact list of WhatsApp user were stored in another database file named as "wa.db". Both the database files can be opened using "sqlite database browser" software which is available for free. The database file "msgstore.db" contains 2 important tables named as "messages" and "chat_list". Table named as "messages" stores all the chat messages in simple text form and the phone numbers with whom chatting was done. Table "messages" was also storing the "Timestamps" which described the time and date of chat sessions. Other table "Chat_list" stores a list of all phone numbers who were added in the users WhatsApp account. The locations of these files were in the folder: **"\data\data\com.whatsapp\databases\"**.Table-10 shows the Artifacts that were found on tested phones.

**Table-11 Artifacts found in the application**

| "Viber" Application Artifacts | Artifacts Found in file "Viber_data" | Artifacts Found in file "Viber_messages" |
|---|---|---|
| | Viber Numbers | 1. Messages to Viber Users in Plain Text |
| | Total number of calls done by user | 2. Phone No.s to whom messages were sent |
| | Phone No.s at which calls were made | 3. Phone No.s from whom messages were received |
| | Duration of Calls to each Phone no. | 4. Date of sent & Received messages |
| | Date of Call | 5. Phone No. with whom conversation took place |
| | | 6. Total number of messages sent to a particular number |

Other Data which was also found in the folder com.whatsapp/files contained the profile picture of the user on Whatsapp account and A Screenshot of Real Phone logs is given at the End of this paper.

## 4.2 Viber Application Forensic Examination & Artifacts
This section describes the forensic analysis of "Viber" Instant Messenger application and Artifacts found during examination of the 5 phones with their results.

### 4.2.1 Artifacts Found in UFED Physical Analyzer
A ".ufd" file generated by UFED device was opened in Physical Analyzer for analysis but **no traces or evidence related to "Viber" was found**.

### 4.2.2 Artifacts Found in Manual Analysis
This section describes the forensic analysis of "Viber" Application and artifacts found on the 5 Android phones with their results. Similar to "WhatsApp", "Viber" Application also maintains and stores a record and logs of all messaging history and calling sessions done and received by the user.
**Screenshot Showing No Viber Artifacts found in Physical Analyzer but showing WhatsApp Artifacts**

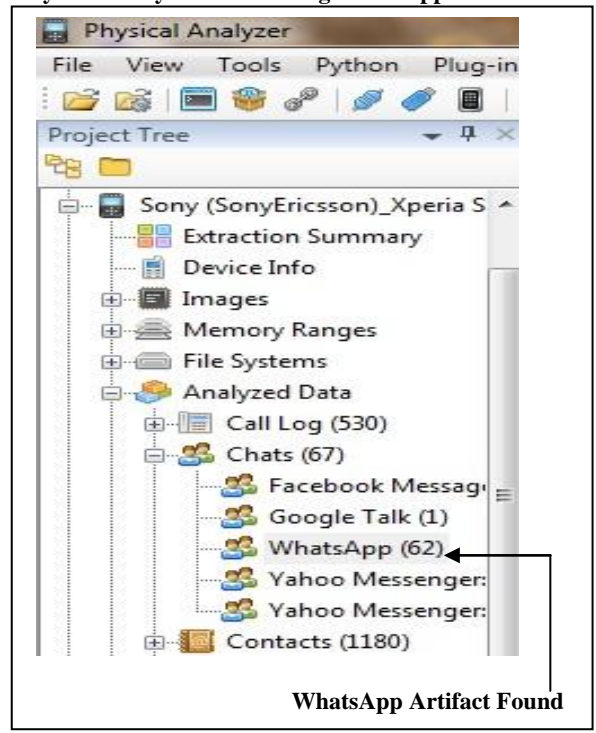

**WhatsApp Artifact Found**

The logs are stored in following 2 different files:
1: Viber_data, 2: Viber_messages
The above 2 files were having no extension to them but it was found that they were database files which can be opened in "Sqlite Database Browser" software, these files were opened and contained very important data, information which is shown in Table-11.

## 5. SUMMARY
In this research, the goal of finding the artifacts related to Instant Messenger applications installed on the phone was achieved which play very important role in any of the real forensic investigations and examinations.
In "WhatsApp" application examination using UFED Physical Analyzer, the chat message artifacts, timestamps and names of files sent and received were found however the storage locations of those files were not found.
In Manual examination of "WhatsApp" application after the FileSystem Extraction, it was found that "msgstore.db" file stores chat list and all chat messages along with timestamps and number's with whom chat sessions were done. "wa.db" file stores all the contacts information. Also, a folder named as "Avatars" in the "com.WhatsApp\files\" folder contains the profile pictures of all the persons with whom chat sessions were done.
In "Viber"application examination using UFED Physical Analyzer, no traces, no artifacts or any message history was obtained.
However, in the Manual examination of folders and files extracted from FileSystem extraction, message logs and call history along with timestamps were obtained. File "Viber_call_log.db" stores information about the calls made by user to other users including phone numbers and timestamps were obtained. Other files "Viber_messages" and "Viber_data" containing sent and received messages were

International Journal of Computer Applications (0975 – 8887)
Volume 68– No.8, April 2013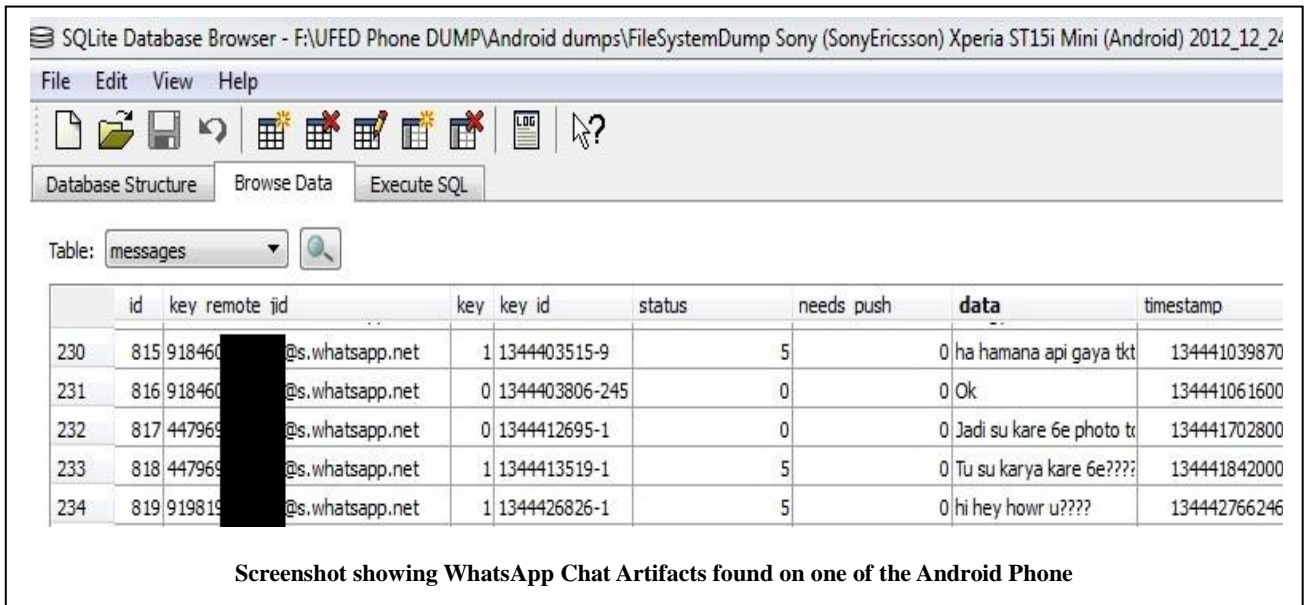

**Screenshot showing WhatsApp Chat Artifacts found on one of the Android Phone**

obtained along with timestamps and contact list was also found. So, rather than using the Physical Analyzer for examination, manual analysis is also important as physical Analyzer didn't displayed any artifacts and data related to "Viber" application.

## 6. CONCLUSION AND FUTURE WORK

This research was aimed to focus on analyzing and finding the artifacts which were live and were present on the Android phone's internal memory and it was achieved. However, Physical Analyzer was able to provide the Artifacts related to "WhatsApp" app only but in manual and folder by folder Analysis of the Extracted data all possible artifacts related to "WhatsApp" and "Viber" applications were found along with TimeStamps. Artifacts included Chat history, messages, received and sent images and video file locations in memory card, contact lists, call details, etc in "Viber". These all artifacts can help forensic investigators and investigation agencies in any cyber investigation. But it was not focused on finding the artifacts if the data is deleted or if the phone is made Factory Reset. Thus, retrieving the artifacts after the factory reset of the phone or retrieving the deleted data can be taken as the future aspect.

## 7. REFERENCES

[1] Alfred Kobsa, Sameer Patil, Bertolt Meyer. (2012). Privacy in Instant Messaging: An Impression Management Model. http://www.ics.uci.edu/~kobsa/papers/2012-B&IT-kobsa.pdf

[2] Mohammad Iftekhar Husain, Ramalingam Sridhar (2010) iForensics: Forensic Analysis of Instant Messaging on Smart Phones http://link.springer.com/chapter/10.1007%2F978-3-642-115349n_2?LI =true#

[3] Android Encryption https://viaforensics.com/category/android-forensics/

[4] Al-Zarouni, Marwan (2006). "Mobile Handset Forensic Evidence: A Challenge for Law Enforcement". http://ro.ecu.edu.au/cgi/viewcontent.cgi?article=1023&context=adf

[5] January 2009, Test Results for Mobile Device Acquisition Tool: Cellebrite UFED 1.1.05 by National Institute of Standards and Technology [NIST]. Available at http://www.ncjrs.gov/pdffiles1/nij/228220.pdf

[6] October 2012, Test Results for Mobile Device Acquisition Tool: CelleBrite UFED 1.1.8.6 -- Report Manager 1.8.3/UFED Physical Analyzer 2.3.0 by National Institute of Standards and Technology [NIST]. Available at http://ncjrs.gov/pdffiles1/nij/238993.pdf

[7] Noora Al Mutawa, Ibrahim Baggili, Andrew Marrington (2012), Digital Investigation (Elsevier), Available at http://www.dfrws.org/2012/proceedings/DFRWS2012-3.pdf

[8] Kailash Kumar, Sanjeev Sofat, S.K.Jain, Naveen Aggarwal (2012). Significance of Hash Value Generation in Digital Forensic: A Case Study. International Journal of Engineering Research and Development. Available at: http://www.ijerd.com/paper/vol2-issue5/I02056470.pdf.

[9] Curran, K., Robinson, A., Peacocke, S., Cassidy, S.(2010) Mobile Phone Forensic Analysis, International Journal of Digital Crime and Forensics, Vol. 2, No. 2, pp:, April-May 2010, ISSN: 1941-6210, IGI Pub

[10] Timothy Vidas, Chengye Zhang, Nicolas Christin (2011). Toward a general collection methodology for Android devices. Available at: http://www.sciencedirect.com/science/article/pii/S1742287611000272

[11] Andre Morum de L. Simao, Fabio Caus Sicoli, Laerte Peotta de Melo, (2011) ACQUISITION OF DIGITAL EVIDENCE IN ANDROID SMARTPHONE. Available at http://igneous.scis.ecu.edu.au/proceedings/2011/adf/9thADFProceedings.pdf#page=12244